# Personal Data Transfers to Non-EEA Domains: A Tool for Citizens and An Analysis on Italian Public Administration Websites

Lorenzo Laudadio
Politecnico di Torino
Turin, TO, Italy
lorenzo.laudadio@polito.it

Antonio Vetrò
Politecnico di Torino
Turin, TO, Italy
antonio.vetro@polito.it

Riccardo Coppola
Politecnico di Torino
Turin, TO, Italy
riccardo.coppola@polito.it

Juan Carlos De Martin
Politecnico di Torino
Turin, TO, Italy
demartin@polito.it

Marco Torchiano
Politecnico di Torino
Turin, TO, Italy
marco.torchiano@polito.it

## ABSTRACT

Six years after the entry into force of the GDPR, European companies and organizations still have difficulties complying with it: the amount of fines issued by the European data protection authorities is continuously increasing. Personal data transfers are no exception. In this work we analyse the personal data transfers from more than 20000 Italian Public Administration (PA) entities to third countries. We developed "Minos", a user-friendly application which allows to navigate the web while recording HTTP requests. Then, we used the back-end of Minos to automate the analysis. We found that about 14% of the PAs websites transferred data out of the European Economic Area (EEA). This number is an underestimation because only visits to the home pages were object of the analysis. The top 3 destinations of the data transfers are Amazon, Google and Fonticons, accounting for about the 70% of the bad requests. The most recurrent services which are the object of the requests are cloud computing services and content delivery networks (CDNs). Our results highlight that, in Italy, a relevant portion of public administrations websites transfers personal data to non EEA countries. In terms of technology policy, these results stress the need for further incentives to improve the PA digital infrastructures. Finally, while working on refinements of Minos, the version here described is openly available on Zenodo: it can be helpful to a variety of actors (citizens, researchers, activists, policy makers) to increase awareness and enlarge the investigation.

## 1 INTRODUCTION

In today's digital and connected world, personal data can be considered a new currency. Commercial data exchanges have a huge impact on the economy, with many companies having built their business around extracting and selling such data [20]. These pieces of information tell much about our lives and contribute to forming what is called our *digital identity*. In the last few decades, several episodes have been reported in which personal data have been collected and exploited without consent from their owners. In 2013, Edward Snowden disclosed the mass surveillance programs conducted by the US government under the name of PRISM [9]. In 2018, Facebook users discovered that their personal data were being collected by the British company Cambridge Analytica and exploited to influence their political opinions [10]. More recently, in 2023, the use of facial recognition systems by Israel to set up biometric surveillance systems targeting Palestinians has been reported by Amnesty International [11].

As a consequence, more and more people seem to have developed concerns about their privacy. At the same time, not everyone seems to be aware of their rights and duties. A 2020 survey by the European Union for Fundamental Rights revealed that 41% of European users are not willing to share any personal data with private companies, but only 22% always read terms and conditions when using online services and only 51% are aware that they can access their own personal data held by organizations [6]. Companies and institutions are not performing very well when it comes to complying with privacy regulations [1].

A 2024 report by DLA Piper showed that since January 2023 the overall amount of fines issued by European supervisory authorities have increased of over 14% with respect to the previous year [17]. In this complex scenario, regulation plays a relevant role. In Europe, the GDPR prohibits personal data transfers from the European Union (EU) and European Economic Area (EEA) to countries for which an adequacy decision is missing and which cannot guarantee appropriate safeguards[2]. In most cases, this is due to the fact that the internal law of those countries conflicts with the Charter of Fundamental Rights of the European Union.

---

[1]See, for example, the highest fines issued for General Data Protection Regulation (GDPR) violations as of January 2024 https://www.statista.com/statistics/1133337/largest-fines-issued-gdpr/ , last visited on 15 May 2024

[2]Even if an adequacy decision is missing and appropriate safeguards cannot be guaranteed from the third country, data transfers can still occur if some exceptions are in place, as we will discuss in the Background section.

In this work, we address the problem of personal data transfers from the EEA to third countries with a focus on data transfers operated by the Italian Public Administration (PA) entities: we provide the justification of our choice in Section 2 together with other contextual information and related work. In Section 3 we report on methodology and instruments of our analysis, presenting Minos: a software tool which is able to detect HTTP requests directed to third countries. We report and discuss the results we obtained by running a mass analysis on the Italian PA entities in Section 4, and identify limitations of the analysis in Section 5. We summarize the contributions and future work in Section 6.

## 2 BACKGROUND AND RELATED WORK

The General Data Protection Regulation (GDPR) is the European regulation which addresses data protection and privacy problems in the EU and EEA, and the transfers of personal data outside the EU and EEA [16]. Article 4 defines personal data as any information which can lead to the identification of a natural person, including location data and online identifiers. The IP address of a host can be considered as personal data. Due to how the Internet works, whenever a personal device makes a request to a certain domain, its IP address is shared with that domain. This means that any HTTP request results in a personal data transfer.

Article 28 states that when the data processing has to be done by processors on behalf of the data controller, the controller should choose only processors which provide sufficient guarantees to implement appropriate technical and organizational measures such that processing meets the requirements of the GDPR and ensures the protection of the rights of the data subject. Also, the data processor should treat personal data only when there is a documented instruction of the controller, even if data is transferred to third countries.

Chapter V of the GDPR regulates data transfers of personal data to third countries or international organizations. Article 45 states that the EC has the power to decide whether a country outside the EU/EEA offers an adequate level of protection for personal data. Such a decision is called an *adequacy decision*, and it implies that personal data can flow freely from the EU and EEA to third countries without any additional protection. If such a decision is missing, data can be transferred if the controller or processor provides appropriate safeguards, enforceable data subject rights and effective legal remedies for data subjects (art. 46). Ultimately, in the absence of adequacy decisions and appropriate safeguards, data transfers can still take place based on exceptions for specific situations, e.g. if the data subject has explicitly consented to the proposed transfer or if the transfer is necessary for important reasons of public interest (art. 49) if appropriate safeguards are provided by the controller or processor, if enforceable data subject rights and effective legal remedies for data subjects are available or if the data subject has explicitly consented to the proposed transfer.

As a main example, we can consider the case of the EU to US data transfers. There is a legal dispute between the EU and the US which has been going on for years, with the approval of the Data Privacy Framework [3] being the latest development. In 2000, the US-EU Safe Harbor Framework allowed EU companies to transfer their data to the US. This framework was invalidated in 2015 by the Court of Justice of the European Union (CJEU), when the Austrian lawyer and privacy advocate Max Schrems filed a complaint against Facebook Ireland Ltd., arguing that the US did not offer adequate protection against surveillance. Schrems' concerns were mainly motivated by FISA 702, a US act which allows for the surveillance of individuals who are not US citizens. In 2016, the European Commission (EC) approved a new agreement called EU-US Privacy Shield, which was later invalidated on the same grounds of the Schrems I judgment. After the abolition of the Data Privacy Shield with the Schrems II judgement, different actors from both the EU and the US stressed the need for a new agreement. Meanwhile, EU organizations have been keeping transferring personal data to the US, while national EU regulators have mostly ignored it. In 2022, the US Department of Commerce and the European Commission developed a new framework called the Data Privacy Framework (DPF). This framework is based on a mechanism which closely resembles the Safe Harbor and Privacy Shield ones: the eligible US-based organizations willing to take part in the DPF program have to self-certify their compliance to certain data protection obligations. On the 10th of July 2023 the European Commission adopted the adequacy decision about the EU-US Data Privacy Framework. New concerns have arisen from Schrems and its non-profit *noyb*, which promised new legal actions [18]. The US has refused to reform FISA 702, which in practice negates reasonable privacy protections to non-US citizens. Moreover, FISA 702 was recently prolonged with modest changes until 2026.

At the beginning of 2024, the European Commission confirmed previous adequacy decisions about 11 non-EU countries, including Israel. Several organizations complained about the inclusion of Israel in this list, writing a letter to the EC to present their concerns [19], claiming that the country's regulations regarding collection and processing of personal data did not comply with the CJEU interpretation of the GDPR standards and the EU Charter of Fundamental Rights (see also Amnesty International's report [11]).

As it can be observed by the regulations and recent developments described above, personal data transfer to non-EEA countries is still an object of public and institutional concern. In our study, we focus on personal data transfers operated by the Italian Public Administration entities because of the highly sensitive nature of the data they operate on. This choice also has another advantage: the Italian PA data are publicly available and updated daily by AGID (Agenzia per l'Italia Digitale, i.e. the Italian Agency for the Development of the National Digital Strategy and Infrastructure).

In addition, to the best of our knowledge, no studies so far addressed the problem of third-countries data transfers concerning the Public Administrations ecosystem. A quantitative investigation on privacy compromising mechanisms on popular websites was conducted by Timothy Libert, with a focus on third-party tracking mechanisms [13]. He found that 88% of the analysed websites made requests to a third-party domain. Guamàn et al. [8] suggested a method to assess the GDPR compliance of data transfers in Android applications, highlighting the presence of ambiguities and inconsistencies within the mobile application environment. They also considered privacy policies for their analysis. They found that 66% of the analysed apps were found as ambiguous, inconsistent or omitting cross-border data transfer disclosures. Granata et al. [7]

---
[3]https://www.dataprivacyframework.gov/Program-Overview

applied the NIST SP-800-53 security control framework to GDPR compliance assessment, introducing a mapping between the GDPR articles and the NIST SP-800-53 security controls. Loré et al. [14] proposed a new AI-based framework to detect potential violations of the GDPR in the context of Italian Public Administration, even though they did not address the analysis of data transfers.

## 3 METHODOLOGY

Our analysis is driven by the following research questions:

- **RQ1:** What is the distribution of Italian PA sources that transfer personal data to non-EEA countries?
- **RQ2:** Which are the most common destinations of the data transfers to non-EEA countries?
- **RQ3:** Which are the most common types of services requested in the context of the data transfers to non-EEA countries?

Herein, we report on the data collection and measurement process.

### 3.1 Data sources

The list of Italian PA entities – along with other information such as the category or official website of each entity – is retrieved from the publicly available OpenData IPA database[4]. The dataset enti.csv contains all the personal data of the entities considered for the analysis: the unique identifier IPA Code[5], Entity name, Category code and Institutional website. The second dataset, categorie-enti.csv, contains all the data about the categories of the entities. The fields of interests are Category name, and Category code, used as key to join this file with the enti.csv file. A new file called entities.csv has been produced from the join of the two datasets, and used as input for the analysis. This latter file contains only the fields of interest.

Some URLs were unusable for the analysis due to being inaccurate, incomplete or inconsistent. Three kind of badly formatted URLs were identified: empty ones, invalid URLs (e.g. about:blank) and those containing typos. Since there is no way to check programmatically whether a given URL is valid or not without contacting the corresponding host, the number of unreachable websites was taken as an estimate of the number of badly formatted URLs. The empty ones were removed before starting the analysis.

### 3.2 Instrumentation

Our investigation required the collection and analysis of a large number of HTTP requests to identify the third countries involved in data transfers with the Italian PA entities.

For this purpose, we developed Minos, a software application that allows non-advanced users to navigate the web while recording their HTTP requests. The tool was developed with the Electron JavaScript framework and is available as open-source on Zenodo [4]. Minos relies upon an internal blacklist which was created by volunteers from the nonprofit community MonitoraPA [6], which operates a public observatory since 2022 and periodically reports about GDPR violations within the Italian Public Administration world. Such blacklist contains popular third-party domains from non-EEA countries[7]. These domains are divided in groups. For example, all the youtube.com.* domains belong to the youtube group. A further refinement has been operated on these domain groups in order to retrieve the companies these domains belong to and their relative countries. For example, both the domains of the microsoft group and the ones of the azure group belong to the US company Microsoft. A mix of different sources was used to trace the companies and countries behind the domain groups, namely: Wikipedia, the website *opencorporates.com* [2] – an open database which contains data of corporations –, the *ipinfo.io* IP Geolocation API service [12], and the *webXray* domain owner list – a large hierarchical collection of third-party domains on the web and the owning companies [3] –.

To automate the analysis of the PA entities, a slightly different version of Minos has been created, called Minos-cli. This version consists of a set of JavaScript and Python scripts to collect, analyse and visualize the data. Minos-cli relies on a running instance of a Chromium-based web browser in order to exploit the Chrome Debugging Protocol (CDP) which allows analysing HTTP requests. A JavaScript module, called Chrome HAR Capturer (CHC), has been modified and used to read data from the fields of the HTTP requests and store them in a csv file [1].

### 3.3 Data collection

The data collection took place according to the following process: (1) the Minos-cli scripts read the list of entities' URLs; (2) a running instance of a Chromium-based web browser is instrumented, in order to contact the website and save the HTTP request in an internal HAR object, through to the Chrome Debugging Protocol; (3a) if an HTTP request's entry.request.url field matches one of the blacklisted domains, a line containing all the data about the request is appended to the bad-requests.csv file, and the IPA Code of the entity being considered is logged to the done.csv file; (3b) if the website cannot be contacted due to some error, the IPA Code of the entity is written to the done.csv file, along with an error message. To detect matches between the request's URL and the blacklisted domains the longest prefix matching is adopted. For example, even if the URL https://www.youtube.com/ matches against both youtube.co and youtube.com, only the latter is taken as valid.

Entities making at least one request to blacklisted domains have been classified as *bad entities*, and their requests as *bad requests*. After data collection, the files bad-requests.csv and done.csv are processed by scripts that compute statistics on the number of bad requests, bad entities, bad entities by category and bad requests by domain, domain group and company.

## 4 RESULTS AND DISCUSSION

Our analysis targeted all the 22890 PA websites present in the OpenData IPA datasets. 14.83% of the websites were not reachable due to errors or missing URLs. The most common error encountered during the analysis was the net::ERR_NAME_NOT_RESOLVED error, which is a symptom of a badly formatted or misspelt URL.

---

[4]https://indicepa.gov.it/ipa-dati/
[5]The names of the fields have been translated to ease the readability for non-Italian speakers
[6]https://monitora-pa.it/

[7]The blacklist can be found at https://github.com/MonitoraPA/Minos/blob/main/hosts.json

### 4.1 RQ1: Transfer of personal data to non-EEA

Figure 1 shows a comparison between the number of good and bad entities. We found that 15% of the entities send data out of EEA countries: since we analyze only the request made to the home page only, we expect this number to be higher in a real navigation context.

Figure 2 displays the top 10 PA categories the bad entities belong to. The category *Municipalities* ranks first: this is consistent with several investigations made in Italy that reported a general trend towards widespread violations of the GDPR by Italian municipalities in recent years. A 2019 report by Federprivacy revealed that at the time 47% of the Italian municipalities did not provide https access to their websites, while 36% did not specify the contact details of their Data Protection Officer, although the regulation has been in force since 2018 [5]. Moreover, between 2020 and 2022, the Italian Data Protection Authority (*Garante per la Protezione dei Dati Personali*) sanctioned several municipalities for unlawful processing or diffusion of citizens' personal data. In general, Italian municipalities seem to have difficulty complying with the GDPR. The second place is held by the category *National teaching institutes*. Several sources have reported an increase in the use by Italian educational institutes of services offered by the US big tech companies in recent years, in particular after the COVID-19 global pandemic. Paolo Monella, in his 2021 article «Education and GAFAM: from awareness to responsibility», reported that since the outburst of the COVID-19 pandemic, schools and colleges based their teaching almost exclusively on big tech infrastructures and platforms [15].

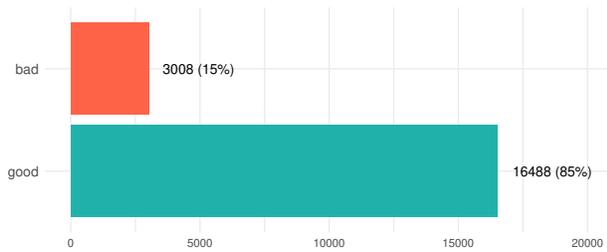

Figure 1: Overall distribution of bad and good entities

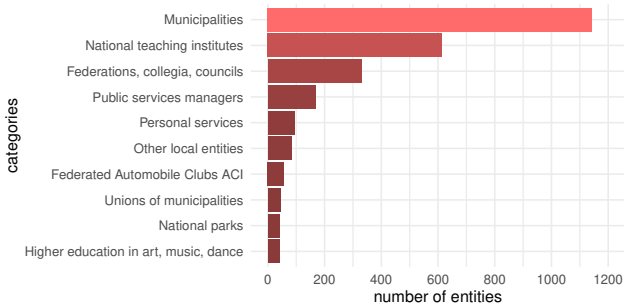

Figure 2: Distribution of bad entities per category

### 4.2 RQ2: destinations of personal data transfers

Figure 3 and Figure 4 visualize the number of bad requests, respectively grouped by target domain group and company.

Most of the requests are addressed to the aws group. Amazon Web Services (AWS) is a subsidiary of Amazon which offers on-demand cloud computing services. Various online sources report AWS as the leading provider of cloud services globally. The website *european-alternatives.eu*[8] lists several cloud service providers from the EU. However, organizations and companies still seem to prefer Amazon for its simplicity, where the self-hosted deployment instead would require advanced knowledge and skills, as well as high deployment costs in the early stages of the operation. Right after, we find Google, another popular big tech giant which offers a wide range of services: maps, web search, website development tools, multimedia hosting, etc. The Google services being more requested are the ones from YouTube, Google Maps and Google Hosted Library (Google's CDN). In third place, we have Fonticons, the company behind Fontawesome, a popular icon library used in many websites. Requests directed to Amazon are more than 40%, and more than 70% of requests are directed to the first three domains (Amazon + Google + Fontawesome). Other companies offer mostly content delivery networks (e.g. Fastly), social networking and multimedia (e.g. Facebook) or other services (e.g. Adobe).

---

[8]https://www.european-alternatives.eu

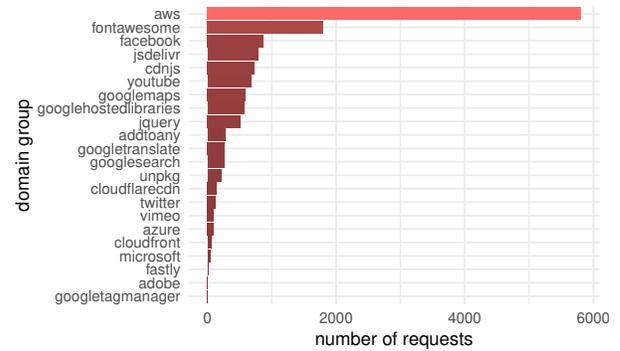

Figure 3: Bad requests per domain group

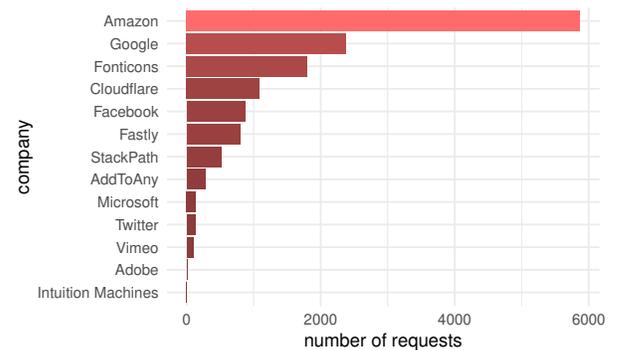

Figure 4: Bad requests per company

## 4.3 RQ3: Distribution type of services

Figure 5 shows the number of bad requests per type of service requested. The results highlight a prevalence of essential services being requested, such as cloud computing services and content delivery networks. Together, the requests for these two types of services represent almost 70% of the overall number of requests. The distribution of requests for cloud computing services is consistent with the global distribution of requests per group: Aws cloud services are the most requested (98%) followed by Azure (2%), Microsoft's cloud services. For CDNs, instead, the distribution is much more uniform: in first place, there is Fontawesome (22%), followed by JSDelivr (21%) and CDNjs (20%). Cloud services and CDNs require a considerable amount of computational power and large IT infrastructures. This is in line with the categories of PA entities identified in RQ1 as most frequently transferring personal data out of EEA: municipalities and teaching institutions, which are - on average - small-sized and, in the absence of a centralized infrastructure provided by the Italian State, do not have the proper know-how and funding to build their own services.

## 5 THREATS TO VALIDITY

The adopted data collection approach produces underestimates of the total number of bad requests. This is due to the fact that when the pages are loaded by the Chromium browser, the session with the website is closed immediately. Therefore, some requests which could be originated by further navigation are not recorded. We are working to improve this aspect and minimize the number of false negatives by waiting for an additional timeout after the page has been loaded and triggering random navigation events with the help of a browser automation tool. While the absolute number of bad requests may be underestimated, we believe that the relative number of bad requests might not be significantly influenced by including navigation to pages other than the considered home pages.

We highlight that the present study considered mainly US destinations of requests (83% of the companies involved in the analysis are from the US). Our goal was to investigate the reliance of Italian PAs on big tech corporations and foreign domains. The majority of domains are from the US simply because many big techs are located in the US.

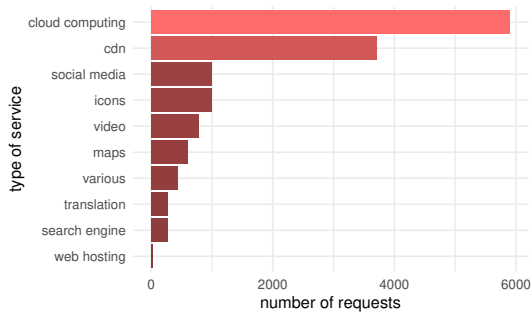

Figure 5: Bad requests per type of service

## 6 CONCLUSION AND FUTURE WORK

In this work, we presented the results of an ongoing analysis of personal data transfers by Italian PA websites to countries outside of the EEA. We leveraged a command line request analyzer, Minos-cli, to run a massive analysis on Italian PA entities from public datasets. Our analysis identified a large number of requests coming from these entities. All of them were directed to US domains, mostly to Amazon domains. The PA entities with the most violations reported were municipalities and schools, requesting most of the time cloud computing services and content delivery networks. Even if some alternatives exist in the EU, numbers are still low, and the self-hosting does not seem a popular alternative, since it requires advanced knowledge and, at least at the beginning, larger investments. In terms of technology policy, these results highlight the need for incentives for the PA's organizations to access IT services within the EEA.

At the moment of writing, we are working to introduce browser automation tools in the data collection phase, performing checks on cookies and checking compliance with the policies of the analysed websites. We also plan to review the lists of domains out of EEA. Therefore, we expect to find a higher quantity of personal data transfers.

In addition, the availability of the Minos tool can enable further investigations by a variety of actors (citizens, activists, policymakers, researchers). The expected long-term contribution of Minos is the increase of awareness about the topic and the empowerment of people by showing which personal data is being transferred and where.

## ACKNOWLEDGMENTS

The authors want to express their sincere gratitude to Giacomo Tesio and Massimo Maria Ghisalberti, who have followed the development of Minos from start to finish, and Marco Ciurcina, whose contribution has been crucial in understanding the legal concepts presented.